\documentclass[notitlepage, 10pt]{revtex4-1}

\usepackage{amsmath}    
\usepackage{graphicx}   
\usepackage{float}
\restylefloat{table}

\begin{document}

\title{Explanation of fractional hierarchy observed experimentally in higher Landau levels}

\author{Janusz Jacak}
\email{janusz.jacak@pwr.edu.pl}
\affiliation{Institute of Physics, Wroc{\l}aw University of Technology, Wyb. Wyspia\'nskiego 27, 50-370 Wroc{\l}aw, Poland}
\author{Lucjan Jacak}
\email{lucjan.jacak@pwr.edu.pl}
\affiliation{Institute of Physics, Wroc{\l}aw University of Technology, Wyb. Wyspia\'nskiego 27, 50-370 Wroc{\l}aw, Poland}

\begin{abstract}
The puzzle of an odd structure of fractional fillings for FQHE in higher Landau levels not repeating the hierarchy from the lowest Landau level is solved. The  fractional filling rates for correlated states in higher Landau levels including spin subbands are systematically derived for the first time. Using topology-type commensurability arguments the hierarchy in higher Landau level fillings is determined in perfect agreement with the experimental observations.   The relative paucity of fractional structure in higher Landau levels is explained and the criterion for pairing in states at half fillings and  for other even-denominator rates  of consecutive   Landau levels is formulated.
\end{abstract}

\pacs{36.40.Gk,73.20.Mf}
\keywords{FQHE hierarchy; higher Landau levels; braid groups; composite fermions; reentrant IQHE}

\maketitle

\section{Introduction} 

The very rich structure of FQHE in the lowest Landau level (LLL) \cite{pan2003} stands in surprising opposition to rather rare manifestation of similar effect in the higher Landau levels (LLs) \cite{lls}. This astonishing observation illustrated widely in experiments with continuous enhancement in precision on still better quality 2DEG samples \cite{lls,ll8/3-1,5/2-1,ll8/3-2,xia2004} had not found an explanation as of yet and is unexpected from the standard composite fermion (CF) interpretation of FQHE \cite{jain,jain2007,hon}. Despite some artificial character, the model of CFs illustrates the so-called Laughlin phase correlations \cite{laughlin1} due to the Aharonov-Bohm phase shift gained when charged particles with attached fluxes mutually interchanges \cite{aharonow-bohm,wilczek} and, moreover, gives rise to an understanding of the main line of hierarchy of fillings factors inside the LLL by mapping of FQHE at this fillings onto IQHE of higher LLs in resultant magnetic field screened by mean field of CF flux-tubes. There are observed, however, many other filling factors even in the LLL for FQHE being outside the main hierarchy, (e.g., $\frac{4}{11}$ or $\frac{5}{13}$), which together with the failure in explanation of correlated states in higher LLs, indicate that not all related Hall state physics is embraced by CFs concept.

\section{Topological approach to planar Hall systems} 

The origin of quantum Hall effects is linked with topology of 2D manifold and can be expressed in terms of homotopy groups of multiparticle configuration space, called braid groups \cite{birman}. In the case of 2D manifold and in the presence of strong magnetic field the special structure of related braid groups emerges referred to cyclotron-braid-subgroups \cite{jac1,ws} of which one dimensional unitary representations (1DURs) weigh nonhomotopic classes of trajectories for path integral \cite{wu,lwitt} in agreement with Laughlin correlations. 
The essence of the construction of cyclotron braids is linked with the interaction of particles (maintaining interparticle separation). The interaction of particles, the flat band condition (assuring the same sized cyclotron orbits for all particles) and the 2D topology (allowing multilooped orbits to reach greater size) constitute necessary prerequisites for the organization of a collective FQH state \cite{ws}.
The cyclotron braid subgroups allow for identification of LL fillings at which the correlated multiparticle state of FQHE can be arranged by virtue of the commensurability condition. In the presence of magnetic field in 2D interacting system, the classical cyclotron orbits may or may not be commensurate with interparticle separation. The commensurability admits definition of braid exchanges of particles and a collective state organization. The method is general and allows for identification of FQHE and IQHE correlations in the system both for the lowest and also for higher LLs. 
The different commensurability of orbits in higher LLs in comparison to LLL directly explains why in higher LLs the FQHE structure is so scanty in comparison to the lowest one, as it will be presented below. 

It has been proved \cite{lwitt} that in the case of not simply-connected configuration space, the path integral formula for the propagator $I_{a\rightarrow b}$ (expressing probability of the system transition form point $a$ to point $b$ in its configuration space) has the form:
$
I_{a\rightarrow b}=\sum\limits_{\eta \in\pi_1} e^{i\alpha_{\eta}}\int d\lambda_{\eta} e^{iS[\lambda_{\eta}(a,b)]/\hbar},
$
$\pi_1$ stands for the full braid group, $S[\lambda_{\eta}(a,b)]$ is the action for trajectory $\lambda_{\eta}(a,b)$ linked $a$ and $b$ with a loop $\eta$ from $\pi_1$, $d\lambda_{\eta}$ is the measure in homotopy sector $\eta$ of trajectories. The factors $ e^{i\alpha_{\eta}}$  form 1DUR of the full braid group \cite{lwitt}. Different representations correspond to different types of quantum particles related to the same classical particles.
For $dimM\geq 3$ ($M$ is a manifold where particles are located) there exist only two 1DURs of the permutation group (i.e., full braid group in this case), corresponding to bosons ($\alpha_{\eta}=0$) and fermions ($\alpha_{\eta}=\pi $), while for $dim M=2$, in particular for $M=R^2$, the infinite number of full braid group 1DURs exists with $\alpha_{\eta}=\alpha \in[0,2 \pi)$ corresponding to anyons.

This, however, does not exhaust all topological resources of 2D configuration space. The specific its property not yet taken into account, manifests itself in the presence of perpendicular magnetic field strong enough to shorten cyclotron trajectories below the interparticle distances. Because classical trajectories must be cyclotron orbits at magnetic field presence, the braid generators, $\sigma_i$---exchanges of neighboring particles, are then excluded as their orbits are too short. Nevertheless, it has been proved \cite{jac1,ws} that exclusively in 2D case multiloop exchanges still match particles but braid generators $\sigma_i$ must be substituted by $(\sigma_i)^q, \; q - odd\; integer$ (note that $(\sigma_i)^q=\sigma_i(\sigma_i)^{q-1}$ and $(\sigma_i)^{q-1}$ makes $\frac{q-1}{2}$ additional loops). The resulting 'cyclotron braid subgroup' generated by generators $(\sigma_i)^q,\; i=1,2,\dots, N$, replaces the full braid group $\pi_1$ in path integral with own 1DURs reproducing phase shift required by Laughlin correlations for LLL filling factor $\nu=\frac{1}{q}$ \cite{jac1,ws}. In this way the Laughlin statistics is obtained without a need to introduce CFs with flux-tubes producing required statistics by Aharonov-Bohm effect.

\begin{figure}[h]
\centering
\includegraphics{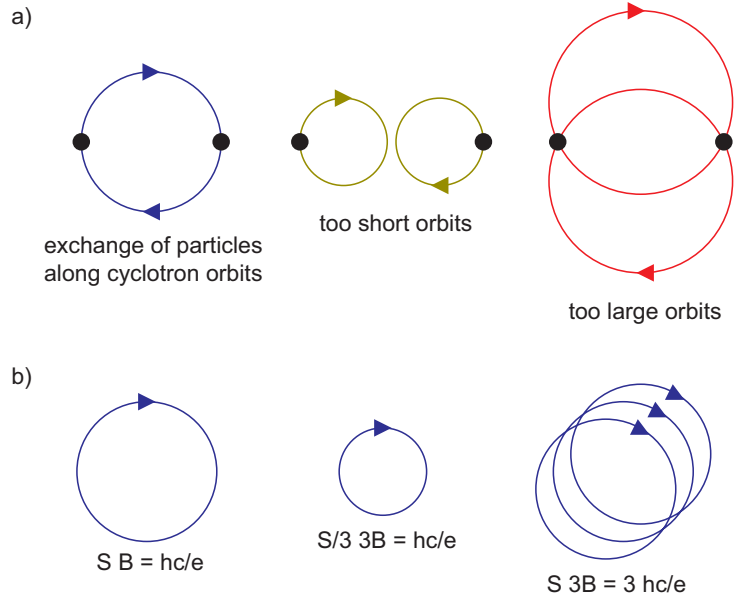}
\caption{(a) Commensurability (blue) of cyclotron orbit with interparticle separation satisfies topology requirements for braid interchanges in equidistantly uniformly distributed 2D particles; for smaller cyclotron radius, particles cannot be matched (brown); for larger the interparticle distance cannot be conserved (red). (b) Schematic illustration of cyclotron orbit enhancement in 2D due to multiloop trajectory structure (third dimension added for visual clarity)}
\label{fig:5}
\end{figure}

To determine LL filling fractions corresponding to FQHE/IQHE collective states we will use thus the condition of commensurability of cyclotron orbits with interparticle 
separation. This condition verifies whether the particle exchange trajectories along cyclotron  orbits, needed for the braid group definition in the presence of the magnetic field, exist or not.  Details of this  approach are described in Refs \citenum{jac1} and \citenum{ws}. For shorthand let us refer here to the illustration in Fig.\ref{fig:5} (a), where the exchange trajectories for two particles are depicted in the topologically  exclusive situation
of ideal  fitting of cyclotron orbits to interparticle distance (left). In the case of too short cyclotron orbits (as e.g., for LLL filling factor $\nu=\frac{1}{3}$) in 2D systems another possibility of exchanges occurs---along multiloop cyclotron orbits---cf. Fig.\ref{fig:5} (b). Because in 2D the surface of a planar orbit must be conserved thus a flux of the external field must be  shared between all loops resulting therefore in reducing of the flux per each loop and in effective increase of orbits. Multiloop structure of classical trajectories described by the cyclotron braid subgroup allows next for determination of quantum statistics, which satisfies Laughlin correlation requirements \cite{jac1}. Thus, even though cyclotron orbit is a classical notion (as needed for  braid groups), its commensurability with particle distribution allows to resolve whether the collective quantum  multiparticle 2D state  can be organized or not. This existential criterion may be applied to determine filling factor structure both of  the lowest and of higher LLs.

The commensurability condition for cyclotron orbits and interparticle separation allows for recovery of   the main FQHE hierarchy in the LLL (similar as within the CF approach), 
\begin{equation}
\label{frac}
\nu =(1)\pm
\frac{l}{l(q-1)\pm 1},
\end{equation} 
here, $q$ is the number of loops (must be odd in order to assure particle interchanges along braids $(\sigma_i)^q$) \cite{jac1,ws}, and $l 
=1,2\dots$  for the main line of the hierarchy \cite{uwaga}, $'1-'$ at the beginning of the r.h.s. of the above expression  corresponds to holes, whereas $\pm$ in the  denominator corresponds to consistent or opposite (figure eight-shaped) orientation of the last loop with respect to others \cite{ws}. The way to derive (\ref{frac}) resolves itself to the decomposition of the external field flux portion per particle when passing through the multiloop cyclotron orbit. The last loop takes away the residual flux if others take away flux quanta per loop. If this residual flux fits to flux portion per particle in $l$-th LL, then the fraction (\ref{frac}) is obtained.

\section{Commensurability and LL filling factors}
Single-particle energy of LLs including spin has the form,
$E_{n\uparrow(\downarrow)}=\hbar\omega_0\left(n+\frac{1}{2}\right) - (+)g\frac{1}{2}\hbar\omega_0$,
where $\hbar\omega_0= \frac{\hbar eB}{mc}= 2\mu_B B $, $\mu_B=\frac{e\hbar}{2mc}$ is Bohr magneton, $g$--giromagnetic factor.	
Degeneracy of each LLs is the same and depends on the magnetic field, 
$N_0=\frac{BS}{hc/e}$,
where $BS$ is the total flux of magnetic field $B$ trough the sample surface $S$, and $\Phi_0=\frac{hc}{e}$ is the quantum of magnetic field flux. 
Let us assume that the sample (2D) surface $S$ is constant as well as the constant number of electrons in the system $N$ is kept. Only the external magnetic field $B$ (perpendicular to the sample) can be changed resulting in variation of LL degeneracy. 

If $B=B_0$, when $N_0=N$, i.e., $N=\frac{B_0 S}{hc/e}$, then the filling factor $\nu =\frac{N}{N_0}=1$ and we deal with completely filled zeroth LL exhibiting fully developed IQHE for spin-polarized electrons.

For higher magnitudes of the field, $ B>B_0$ we have $N_0>N$, i.e., $\nu=\frac{N}{N_0}<1$. In this case the cyclotron orbits are too short to reach neighboring particles and in order to organize a collective state the multilooped cyclotron orbits are required. For $q$-loop orbits the commensurability condition,
$q\frac{hc}{eB}=\frac{S}{N}$
gives 
$\nu=\frac{N}{N_0}=\frac{SeB/qhc}{SeB/hc}=\frac{1}{q}$.
Because $q$ must be odd to prevent that corresponding exchange trajectories are braids \cite{jac1,ws}, we get in this manner main ratios for FQHE. 
1DUR of cyclotron braid subgroup gives the phase shift when particle exchange. This particle exchange is understand as an exchange of arguments of the wave function and this function must gain then the phase shift according to 1DUR of braid describing this exchange \cite{imbo1988,sud}, 
in perfect coincidence with Laughlin correlations. 

Now let us consider $B<B_0$, that $2N_0>N>N_0$. Then the subband $0\uparrow$ is completely filled with $N_0$ electrons, while the rest of electrons, $N-N_0<N_0$ must be located in $0\downarrow$ subband. For electrons $0\uparrow$ their cyclotron orbits equal to $\frac{hc}{eB}=\frac{S}{N_0}$ and these electrons form a collective state of IQHE. The electrons $0\downarrow$ fill the subband $0\downarrow$ and their cyclotron orbits $\frac{hc}{eB}$ are certainly shorter in comparison to particle separation $\frac{S}{N-N_0}$. The multiloop trajectories are thus needed corresponding to FQHE for these electrons.
For $q$-loop ($q$ is an odd integer to assure particle exchanges),
$q\frac{hc}{eB}=\frac{S}{N-N_0}$,
thus $\nu=\frac{N}{N_0}=1+\frac{1}{q}=\frac{4}{3}, \frac{6}{5},\frac{8}{7},\dots$
and dually for holes in this subband, $\nu =2-\frac{1}{q}=\frac{5}{3},\frac{9}{3}, \frac{13}{7}, \frac{17}{9},\dots$. In the subbband $0\downarrow$ the filling hierarchy structure is repeated from the $0\uparrow$ subband. 

For lower field $B$, that $N=2N_0$, the cyclotron orbits in both subbands $0\uparrow$ and $0\downarrow$ ideally fit to interparticle separations of two sorts of particles $0\uparrow$ and $0\downarrow$, i.e., $\frac{hc}{eB}=\frac{S}{N_0}=\frac{S}{N/2}$, and in both subbands of LLL we deal with IQHE.

If now the magnetic field is still lowering, that $3N_0>N>2N_0$, then the subbands of LLL, $0\uparrow$ and $0\downarrow$, are completely filled with IQHE for electrons in these subbands, while the rest of electrons, i.e., $N-2N_0$ is located in the subbband $1\uparrow$.
There are now the following possibilities:
The cyclotron orbit from this subband equals to $\frac{3hc}{eB}$ (note that the LL energy without the spin contribution, i.e., the kinetic energy of flat band decides on the size of cyclotron radius at field $B$) and radius may ideally fit to interparticle separation in this subband, then 
$\frac{3hc}{eB}=\frac{S}{N-2N_0}$,
thus $\nu=\frac{N}{N_0}=\frac{7}{3}$ because $\frac{3hc}{eB}=\frac{3S}{N_0}=\frac{S}{N-2N_0}$, and in all subbands we deal with IQHE (though the subband $1\uparrow$ is not completely filled).
The another situation is when the cyclotron orbit $\frac{3hc}{eB}$ in this band is too short to match particles in this subband, $ \frac{3hc}{eB}<\frac{S}{N-2N_0}$. Then the multiloop ($q$-loop, $q$ odd integer) trajectories are needed resulting then in FQHE for these electrons: 
$q\frac{3hc}{eB}=\frac{S}{N-2N_0}$,
then $\nu = \frac{6q+1}{3q}=\frac{19}{9}, \frac{31}{15}, \frac{43}{21},\dots $. 
The next possibility for commensurability of cyclotron orbits and interparticle separation of electrons from the subband $1\uparrow$ is in the case when the cyclotron orbit is twice in dimension of interparticle separation:
$\frac{3hc}{eB}=\frac{2S}{N-2N_0}$,
then $\nu=\frac{8}{3}$, because in this case $\frac{3S}{N_0}=\frac{2S}{N-2N_0}$. In this case we deal with IQHE in all subbands including the last one incompletely filled, however. 
For lower field $B$, when the cyclotron orbit in the subband $1\uparrow$,
$\frac{3hc}{eB}$, fits to three interparticle separations, $\frac{S}{N-2N_0}$, i.e.,
$\frac{3hc}{eB}=\frac{3S}{N-2N_0}$, 
then $N=3N_0$, and $\nu = \frac{N}{N_0}=3$, which corresponds to completely filled three first subbands, $0\uparrow$, $0\downarrow$ and $1\uparrow$, with IQHE correlation in each subband.

When we still lower the magnetic field $B$ we attain the region $4N_0>N>3N_0$, corresponding to filling of $1\downarrow$ subband. In this case three antecedent subbands $0\uparrow,\;0\downarrow,\;1\uparrow$ are completely filled. In subbands $0\uparrow,\;0\downarrow$ cyclotron orbits are of size $\frac{hc}{eB}$ and they are ideally commensurate with interparticle separation in these subbands equal to $\frac{S}{N_0}$. Thus the correlation in these subbands corresponds to IQHE.
In the subband $1\uparrow$ the cyclotron orbit has a surface $\frac{3hc}{eB}$ three times larger than $\frac{S}{N_0}$. The commensurability condition allows thus the exchanges of every third electron along singleloop orbits, i.e., IQHE on such subset of electrons. 
In the subband $1\downarrow$ we have $N-3N_0<N_0$ electrons. The cyclotron orbit $\frac{3hc}{eB}$ may be here too small, equal or too large in comparison to $\frac{S}{N-3N_0}$.
If this cyclotron orbit is too small for interchanges, then multiloop trajectories are required resulting in FQHE of related electrons. This happens at $\nu=\frac{N}{N_0}=\frac{9q+1}{3q}=\frac{28}{9},\frac{46}{15}, \frac{64}{21},\dots$ (because $\frac{3qS}{eB}=\frac{S}{N-3N_0}$, $q$-odd). In this case the correlation in $1\downarrow$ subband has FQHE character.
If $\frac{3hc}{eB}=\frac {3S}{N_0}=\frac{S}{N-3N_0}$, then $\nu =\frac{10}{3}$ and the corresponding correlation in $1\downarrow$ has IQHE character.
For $\frac{3hc}{eB}=\frac{3S}{N_0}=\frac{2S}{N-3N_0}$, i.e., for $\nu=\frac{N}{N_0}=\frac{11}{3}$, the commensurability holds for every second electron in the subband resulting in IQHE correlation for electron subset in this subband. 
Finally, for $\frac{3hc}{eB}=\frac{3S}{N_0}=\frac{3S}{N-3N_0}$ we have $\nu=4$ and IQHE corresponding to commensurability of cyclotron orbit with every third electron in the subband $1\downarrow$.
The symmetric ratios for holes in the subband $1\downarrow$ must be also taken into account. 

For sufficiently low magnetic field, that $5N_0>N>4N_0$, i.e., $\nu \in (4,5]$ we deal with four first LL subbands completely filled. In subbands $0\uparrow$ and $0\downarrow$ the cyclotron orbits $\frac{hc}{eB}$ fit to interparticle separation, i.e., $\frac{hc}{eB}=\frac{S}{N_0}$. In subbands $1\uparrow$ and $1\downarrow$ the commensurability condition has the form $\frac{3hc}{eB}=\frac{3S}{N_0}$. 
The remaining electrons fill now the subband $2\uparrow$.
In the subband $2\uparrow$ there are $N-4N_0<N_0$ electrons and cyclotron orbits have the size $\frac{5hc}{eB}$.
If cyclotron orbits in $2\uparrow$ fit to interparticle separation, 
$\frac{5hc}{eB}=\frac{S}{N-4N_0}$,
then $\nu=\frac{21}{5}$ corresponding to IQHE in all filled subbands and fractionally filled $2\uparrow$ subband.
For cyclotron orbit shorter than separation of particles in the subband $2\uparrow$ the multiloop cyclotron orbits are required resulting in FQHE in this subband. This happens for filling ratios: 
$\frac{q5hc}{eB}=\frac{S}{N-4N_0}
=\frac{p5S}{N_0} \rightarrow 
\nu=\frac{20q+1}{5q}=\frac{61}{15},\frac{101}{25},\frac{141}{35},\dots$,
corresponding to FQHE in the subband $2\uparrow$.
For commensurability condition $\frac{5hc}{eB}=\frac{2S}{N-4N_0}$, $\nu=\frac{22}{5}$, corresponding to IQHE in the $2\uparrow$ subband. Similarly for $\frac{5hc}{eB}=\frac{iS}{N-4N_0},\;i=3,4$, $\nu=\frac{23}{5},\frac{24}{5}$.
Finally, for $\frac{5hc}{eB}=\frac{5S}{N-4N_0}$, $\nu=5$ and we deal with IQHE of completely filled first five LL subbands.

The corresponding filling ratios are summarized in Table \ref{tab1} for five first subbands of LL structure, cf. also Fig. \ref{fig:1} where pushing of FQHE toward the edge of band in higher LLs is noticeable. The generalization of hierarchy (\ref{frac}) for the $n$th LL  attains the form,
 \begin{equation}
\label{frac1}
\nu=\left\{ \begin{array}{l}
2n(+1)\pm\frac{l}{l(2n+1)(q-1)\pm 1},\;for\;\uparrow,\\
2n+1(+1)\pm\frac{l}{l(2n+1)(q-1)\pm 1},\;for\;\downarrow,\\ \end{array}\right.
\end{equation} 
where $n$ is the LL's number.

\section{Even-denominator fractions}
The commensurability conditions allow also for systematic analysis of some special fractional fillings of LLs expressed by ratios with even denominators. The most prominent seem to be the ratios exclusively in $n>0$ LLs of the form, $\nu =\frac{5}{2}, \frac{7}{2}, \frac{9}{2}, \frac{11}{2} \dots$, resulting from special commensurability condition, 
\begin{equation}
\label{frac3}
\frac{(2n+1)hc}{eB}=\frac{\frac{2n+1}{2}S}{N-\left\{
\begin{array}{ll}
2n N_0 & for\;\uparrow \\
(2n+1)N_0 &for \;\downarrow
\end{array}\right.},
\end{equation}
for subbands $n\uparrow (\downarrow)$, respectively. By pairing of particles in the last subband the ideal commensurability can be achieved for the pairs allowing IQH ordering for them (this is due to twice reducing of denominator in r.h.s. of the above formula for pairs, while the cyclotron orbits conserve their size for pairs because cyclotron radius scales as $\sim\frac{e}{m}=\frac{2e}{2m}$). This happens for $\frac{5}{2},\frac{7}{2},\frac{9}{2}, \dots$, but not for $\nu=\frac{1}{2}$ and $\frac{3}{2}$ from $n=0$ spin subbands, where cyclotron orbits always are shorter than separation of particles, what precludes pairing. The latter two ratios correspond to Hall metal states, the condition for which can be obtained from formulae (\ref{frac}) and (\ref{frac1}) taken in the limit $l\rightarrow\infty$. These properties remarkably agree with experimental observations and numerical analysis of states proposed for related filling ratios as it has been summarized recently in Ref. \citenum{5/2-1}.

\begin{table}[th]
\centering
\begin{tabular}{|p{1.8cm}|p{5.5cm}|p{6.5cm}|p{3.5cm}|}
\hline
LL subb.&IQHE&FQHE ($q-odd,\; l=1,2,3,\dots)$
&Hall metal \\
\hline
$0\uparrow$& 1 & 
$\frac{1}{q}, 1-\frac{1}{q}, \frac{l}{l(q-1)\pm1},1-\frac{l}{l(q-1)\pm 1}$&$\frac{1}{q-1},
1-\frac{1}{q-1}$ \\
\hline
$0\downarrow$&2& $1+\frac{1}{q}$,
$2-\frac{1}{q}, 1+\frac{l}{l(q-1)\pm 1}, 2-\frac{l}{l(q-1)\pm 1}$& $1+\frac{1}{q-1},
2-\frac{1}{q-1} $
\\
\hline
$1\uparrow$&$\frac{7}{3},\frac{8}{3},3$,
($\frac{5}{2}\; paired$)& $2+\frac{1}{3q},2+\frac{l}{3l(q-1)\pm 1},3-\frac{1}{3q}, 3-\frac{l}{3l(q-1)\pm 1} $& $2+\frac{1}{3(q-1)},3-\frac{1}{3(q-1)}$ \\
\hline
$1\downarrow$&$\frac{10}{3},\frac{11}{3},4$,($\frac{7}{2}\; paired$) &$3+\frac{1}{3q}, 3+\frac{l}{3l(q-1)\pm 1}, 4-\frac{1}{3q},4-\frac{l}{3l(q-1)\pm1}$&
$3+\frac{1}{3(q-1)}, 4-\frac{1}{3(q-1)}$\\
\hline
$2\uparrow$&$\frac{21}{5},\frac{22}{5},\frac{23}{5},\frac{24}{5},5$,($\frac{9}{2}\; paired$) &
$4+\frac{1}{5q}, 4+\frac{l}{5l(q-1)\pm 1},5-\frac{1}{5q}, 5-\frac{l}{5l(q-1)\pm 1}$&$ 4+\frac{1}{5(q-1)}, 5-\frac{1}{5(q-1)}$\\
\hline
$2\downarrow$&$\frac{26}{5}, \frac{27}{5}, \frac{28}{5}, \frac{29}{5},6$,($\frac{11}{2}\; paired$) &
$5+\frac{1}{5q}, 5+\frac{l}{5l(q-1)\pm 1},
6-\frac{1}{5q}, 6-\frac{l}{5l(q-1)\pm 1}$&$5+\frac{1}{5(q-1)}, 6-\frac{1}{5(q-1)}$\\
\hline
$3\uparrow$&$\frac{43}{7}, \frac{44}{7}, \frac{45}{7}, \frac{46}{7}, \frac{47}{7}, \frac{48}{7}, 7$,($\frac{13}{2}\; paired$) &
$6+\frac{1}{7q}, 6+\frac{l}{7l(q-1)\pm 1},
7-\frac{1}{7q},7-\frac{l}{7l(q-1)\pm 1}$&$ 6+\frac{1}{7(q-1)}, 7-\frac{1}{7(q-1)}$\\
\hline
$3\downarrow$&$\frac{50}{7}, \frac{51}{7}, \frac{52}{7}, \frac{53}{7}, \frac{54}{7}, \frac{55}{7}, 8$,($\frac{15}{2}\; paired$) &
$7+\frac{1}{7q},
7+\frac{l}{7l(q-1)\pm 1},8-\frac{1}{7q},8-\frac{l}{7l(q-1)\pm 1}$&
$7+\frac{1}{7(q-1)}, 8-\frac{1}{7(q-1)}$\\
\hline
\end{tabular}
\caption{LL filling factors for IQHE and FQHE (main) determined by commensurability arguments ($paired$ indicates condensate of electron pairs);  for $l=\frac{4}{3},\frac{5}{3},\frac{3}{2}\dots$ one can obtain in subband $0\uparrow$ all rates out of the main line, cf.  \cite{uwaga} (note that $\pm $ in the denominators can be  formally substituted by $l \rightarrow \pm l $)}
\label{tab1} 
\end{table}

\section{Comparison with experimental observations}
\begin{figure}[h]
\centering
{\includegraphics{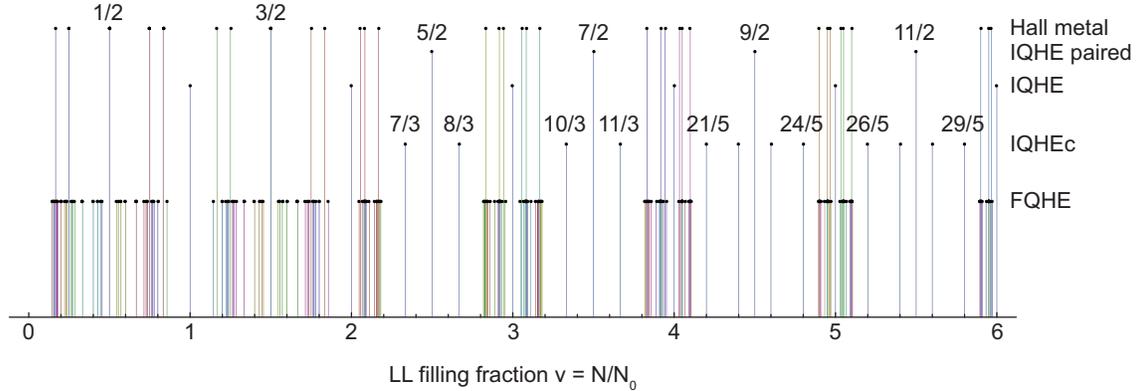}}
\caption{Graphical presentation of filling factors up to sixth LL subband selected by commensurability condition: each spike represents filling ratio for possible correlated state, the lowest ones correspond to series of FQHE hierarchy given by (\ref{frac}) and (\ref{frac1}), next to IQHE inside bands with $ n\geq 1$, the tallest ones correspond to paired states acc. (\ref{frac3}) and Hall metal states acc. (\ref{frac}) and (\ref{frac1}) taken in the limit $ l\rightarrow \infty$;    the evolution of FQHE fillings is visible with growing number of LL subband; IQHE c indicates states selected by commensurability condition corresponding to single cyclotron orbits as in completely filled subband but not protected by large inter-level energy gap; IQHE paired indicates states with pairing preferable by commensurability}
\label{fig:1}
\end{figure}

In Ref. \citenum{lls} the odd structure of FQHE for $n=1,2$ in comparison to $n=0$ LL is reported. In ultra-low temperatures $\sim 15$mK and in of high mobility samples of GaAs/AlGaAs ($\mu\sim 11\times 10^6$cm$^2$/Vs), the pronounced correlated state features at $\nu=\frac{7}{3}$ and $\frac{8}{3}$ were observed in magneto-transport experiment. They were interpreted as FQHE accompanied also such ordering at $\nu=\frac{5}{2}$. It remains in contrast with the plethora of FQHE in the lowest LL. The relative absence of FQHE in higher LLs well coincides with the predictions based on commensurability conditions (Fig. \ref{fig:1}). From this approach it follows that the states $\frac{7}{3},\frac{8}{3}$ would be rather of IQHE c type and not related with multiloop trajectories typical for FQHE. 
The nature of states $\frac{7}{3},\frac{8}{3}$ has been widely investigated both theoretically and experimentally \cite{ll8/3-1,ll8/3-2}. These states are linked to $\frac{5}{2}$ state regarded as FQHE for filling fraction with even denominator, though pairing expressed by various variants of Pfaffian function was also taken into account for this state \cite{5/2-1}. Related to postulated fractional type of correlation for these states the possibility of non-Abelian charges convenient for topological quantum information processing was considered \cite{5/2-1,ll8/3-2,ll8/3-1}. Thus suggestion that states $\frac{7}{3} , \frac{8}{3} $ are not of FQHE type but rather of IQHE c type would contribute this discussion.

\begin{figure}[h]
\centering
\includegraphics{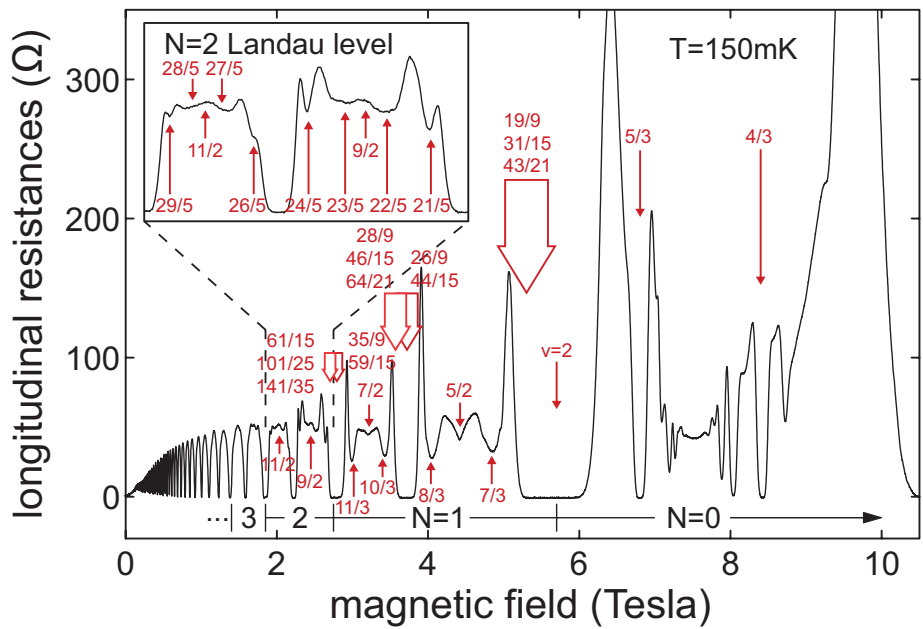}
\caption{Resistivity measurements for wide range of magnetic field corresponding to $n=1,2$ in high mobility GaAs/AlGaAs heterostructure (after Ref. \citenum{lls}), in color  are indicated fractions predicted by the commensurability arguments (Fig. \ref{fig:1})  in perfect agreement with the experimentally observed features}
\label{fig:9}
\end{figure}

In $n=2$ LL situation again changes \cite{lls} and two states with even denominator $\frac{9}{2}$ and $\frac{11}{2}$ occurred revealing, however, strong anisotropy in comparison to non-anisotropic states at $\frac{5}{2}$ and $\frac{7}{2}$. These all states are accompanied by local minima, which perfectly fit to IQHE c states predicted by commensurability criterion, $\{\frac{26}{5},\frac{27}{5},\frac{28}{5},\frac{29}{5}\}$,
$\{\frac{21}{5},\frac{22}{5},\frac{23}{5},\frac{24}{5}\}$, $\{\frac{10}{3}, \frac{11}{3}\}$ and $\{\frac{7}{3},\frac{8}{3}\}$, for IQHE paired states, $\frac{11}{2}$, $\frac{9}{2}$, 
$\frac{7}{2}$ and $\frac{5}{2}$, respectively (cf. Table \ref{tab1}, Figs \ref{fig:1} and \ref{fig:9}). 

Another feature predicted by commensurability conditions, i.e., location of series of FQHE states in higher Landau LLs ($n\geq 1$) closely to the subband edges also seems to be consistent with experimental observations, as it is marked in Fig. \ref{fig:9} by thick arrows. Densely lying FQH states in vicinity of IQH state are probably not distinguishable due to resolution reason and mix together resulting in larger flattening of main IQHE minimum in resistivity measurement. 

\section{Conclusion}
The topological braid group-based approach to FQHE/IQHE turns out to be effective in recognition of fractional higher LL fillings and in explanation of distinct structure of FQHE in these levels in comparison to the LLL case.
The reason of this property is identified by cyclotron braid group treatment in different  commensurability of cyclotron orbits and interparticle separation in various LLs.
With the growing number of occupied LL, the corresponding cyclotron orbit becomes larger because of the growing kinetic energy and thus an exceeding of the orbital size by interparticle distance, the effect critical for FQHE, occurs in higher subbands at relatively small densities of particles only, close to subband edges.
Thus FQHE is gradually pushed toward the edge of the subband with growing subband number in opposition to the LLL, inside which cyclotron orbits were always too short for particle interchanges (in braid group terms). Simultaneously, in higher LLs some new commensurability opportunities occur which were impossible in the LLL. This new property resembles commensurability in completely filled subbands and corresponds to singleloop orbits characteristic for IQHE, but without the large energy gap protecting these special fillings as for completely filled LL subbands. The filling ratios selected by this type of commensurability are visible in experiment similarly as also predicted paired states in centers of subbands for $n\geq 1$.  For the latter  case, the not affected by pairing cyclotron orbits are commensurate with the enhanced separation between pairs in comparison to single particles for $n\geq 1$, but not for $n=0$, where an increase of separation only worsens  an opportunity to fit with too short cyclotron radius. Therefore, in the LLL at half filling ($\nu=\frac{1}{2}, \frac{3}{2}$) we deal  with unpaired Hall metal, whereas for half fillings of higher LLs, the pairing is necessary. The Hall metal states in higher LLs occur, however, in association to fractional series close to band edges, if one takes the limit $l\rightarrow \infty$ in the generalized hierarchy (\ref{frac1}), i.e., for zero residual flux fraction per the last loop of the multiloop orbit. These predictions based on commensurability of cyclotron orbits with particle distribution perfectly agree with experimental observations and with numerical  exact diagonalization studies. 

\begin{acknowledgments}
Authors acknowledge the support of the present work upon the NCN project no. 2011/02/A/ST3/00116.
\end{acknowledgments}


\begin{thebibliography}{10}
\bibitem{uwaga}
for '-' in the denominator in Eq. (\ref{frac}), $l$=2\dots; moreover, for
  not-integer $l$, but for some characteristic fillings above the first band,
  e.g., $l=\frac{4}{3},\frac{5}{3},\frac{3}{2}$, where $l$ displays in
  (\ref{frac}) the filling fraction accommodated to the last loop of multiloop
  cyclotron trajectory, one obtains acc. to the concept of commensurability,
  the fractions $\nu$ out of the main line,
  $\nu=\frac{4}{11},\frac{4}{5},\frac{5}{13},\frac{5}{7},\frac{3}{8},\frac{3}{10}\dots$.
\end{thebibliography}
\end{document}